\documentclass[11pt,preprint]{aastex}

\slugcomment{To appear in the Astrophysical Journal, 20 June 2005\\
{\it submitted 3 February 2005; accepted 1 March 2005}}

\shorttitle{Thermal Emission from an Extrasolar Planet}
\shortauthors{Charbonneau et al.}

\begin{document}

\title{Detection of Thermal Emission from an Extrasolar Planet}

\author{David Charbonneau\altaffilmark{1,2}, Lori E. Allen\altaffilmark{1}, 
S. Thomas Megeath\altaffilmark{1}, Guillermo Torres\altaffilmark{1}, 
Roi Alonso\altaffilmark{3}, Timothy M. Brown\altaffilmark{4}, Ronald L. Gilliland\altaffilmark{5}, 
David W. Latham\altaffilmark{1}, Georgi Mandushev\altaffilmark{6}, 
Francis T. O'Donovan\altaffilmark{7}, and Alessandro Sozzetti\altaffilmark{8,1}}
\altaffiltext{1}{Harvard-Smithsonian Center for Astrophysics, 60 Garden St., Cambridge, MA 02138}
\altaffiltext{2}{dcharbonneau@cfa.harvard.edu}
\altaffiltext{3}{Instituto de Astrof{\'i}sica de Canarias, 38200 La Laguna, Tenerife, Spain}
\altaffiltext{4}{High Altitude Observatory, National Center for Atmospheric Research, 3450 Mitchell Lane, Boulder, CO 80307}
\altaffiltext{5}{Space Telescope Science Institute, 3700 San Martin Dr., Baltimore, MD 21218}
\altaffiltext{6}{Lowell Observatory, 1400 W. Mars Hill Rd., Flagstaff, AZ 86001}
\altaffiltext{7}{California Institute of Technology, 1200 E. California Blvd, Pasadena, CA 91125}
\altaffiltext{8}{Department of Physics and Astronomy, University of Pittsburgh,
Pittsburgh, PA 15260}

\begin{abstract}
We present {\it Spitzer Space Telescope} infrared photometric time series of the transiting 
extrasolar planet system TrES-1.  The data span a predicted time of secondary eclipse,
corresponding to the passage of the planet behind the star.  In both
bands of our observations, we detect a flux decrement with a timing, amplitude, and 
duration as predicted by published parameters of the system.  This signal represents
the first direct detection of (i.e. the observation of photons emitted by) 
a planet orbiting another star.  The observed eclipse depths (in units of relative
flux) are $0.00066\pm0.00013$ at 4.5~$\mu$m and $0.00225\pm0.00036$ at 8.0~$\mu$m.
These estimates provide the first observational constraints on models of the thermal emission of 
hot Jupiters.  Assuming that the planet emits as a blackbody, 
we estimate an effective temperature of $T_{p} = 1060 \pm 50$~K.
Under the additional assumptions that the planet is in thermal
equilibrium with the radiation from the star and emits isotropically, 
we find a Bond albedo of $A = 0.31 \pm 0.14$.  This would imply that the planet 
absorbs the majority of stellar radiation incident upon it, a conclusion
of significant impact to atmospheric models of these objects. 
We also compare our data to a previously-published model of the planetary 
thermal emission, which predicts prominent spectral features
in our observational bands due to water and carbon monoxide.  This model 
adequately reproduces the observed planet-to-star flux ratio at 8.0~$\mu$m, however
it significantly over-predicts the ratio at 4.5~$\mu$m.
We also present an estimate of the timing of the secondary eclipse,
which we use to place a strong constraint
on the expression $e \, \cos{\omega}$, where $e$ is the orbital eccentricity 
and $\omega$ is the longitude of periastron.  
The resulting upper limit on $e$ is sufficiently small that we conclude 
that tidal dissipation is unlikely to provide a significant source of energy
interior to the planet.
\end{abstract}

\keywords{binaries: eclipsing --- infrared: stars --- planetary systems --- stars: individual (\objectname[GSC~02652-01324]{TrES-1}) --- techniques: photometric}

\section{Introduction}
Extrasolar planets that transit their parent stars are particularly
valuable since they afford direct estimates of key physical
parameters of the object.  Moreover, it is only for bright stars
that direct follow-up studies of the planet are likely to succeed:
Indeed, the numerous follow-up observations of the brightest-known transiting
system HD~209458 \citep{charbonneau2000, henry2000} include 
the detection \citep{charbonneau2002} and upper limits \citep{richardson2003a,
richardson2003b, deming2005a} 
of absorption features in the planetary atmosphere, 
the discovery of a cloud of escaping hydrogen atoms \citep{vidal2003}, 
and a search for circumplanetary rings and Earth-sized satellites 
\citep{brown2001}.
The OGLE survey \citep{udalski2002a, udalski2002b, udalski2003, udalski2004} 
has yielded 5 transiting planets 
\citep{bouchy2004, konacki2003, konacki2005, moutou2004, pont2004}
with reasonably precise estimates
of the planetary masses and radii; however, the direct follow-up studies
described above are impeded for these objects due to their great distance
and resulting faintness.  
The exclusive opportunities afforded by bright systems 
have provided strong motivation for numerous wide-field,
small-aperture surveys for these objects 
\citep{alonso2004, bakos2004, christian2004, dunham2004, odonovan2004, pepper2004}.
The first success for the wide-field approach occurred
only recently with the detection of TrES-1 \citep{alonso2004}. 
This hot Jupiter has a mass $M_{p} = 0.76 \pm 0.05 \, M_{\rm Jup}$ 
and radius $R_{p} = 1.04^{+0.08}_{-0.05} \, R_{\rm Jup}$, and is 
located 0.0394~AU from the central K0V star \citep{sozzetti2004}.  

Soon after the discovery of TrES-1,
we proposed for Director's Discretionary time on the 
{\it Spitzer Space Telescope} to monitor photometrically the TrES-1 system
with the goal of detecting the thermal emission of the planet
through observation of the secondary eclipse (i.e. the passage
of the planet behind the star).  Detection of this signal
is of interest for several reasons:  First, it would constitute
the first direct detection of an extrasolar planet, i.e.
the detection of photons emitted by the planet.  Second, the amplitude
of the secondary eclipse can be compared directly to theoretical
models of the planetary atmosphere and, under the assumption
of blackbody emission, yields an estimate of the effective
temperature of the planet.  Third, the timing and duration
of the secondary eclipse place constraints on the orbital
eccentricity that are much more restrictive than those from
the radial velocity orbit alone.  A non-zero eccentricity requires a source
of excitation (such as a second planet), and the damping of this
orbital eccentricity provides an internal energy source that
could slow the contraction of the radius of the transiting planet \citep{bodenheimer2001, bodenheimer2003}.
Conversely, a zero eccentricity would rule out this scenario.

In this paper, we present the detection of the thermal
emission from the extrasolar planet TrES-1, and discuss the 
implications for our understanding of the planet. 

\section{Observations and Time Series Production}
The Infrared Array Camera \citep[IRAC;][]{fazio2004}
on the {\it Spitzer Space Telescope} obtains simultaneous images 
in four bandpasses.  A $5\farcm2 \times 5\farcm2$ field of view  (FOV)
is imaged in one pair of bandpasses (3.6~$\mu$m \& 5.8~$\mu$m), and a 
nearly adjacent $5\farcm2 \times 5\farcm2$ FOV is imaged in the second pair 
(4.5~$\mu$m \& 8.0~$\mu$m).  The two blue channels employ InSb detectors,
whereas the red channels use Si:As IBC detectors.  All four arrays are 
$256\times256$ pixels.

We elected to monitor TrES-1 in only one channel pair (4.5~$\mu$m 
\& 8.0~$\mu$m) to avoid repointing the telescope
during the course of the observations.  We chose
not to dither the pointing in order to minimize the motion of
the stars on the pixel array.  We carefully selected the
pointing position so that TrES-1 (2MASSJ19040985+3637574; $J=10.294$, 
$J-K$=0.475) and two bright calibrators
(2MASSJ19041058+3638409; $J=9.821$, $J-K=0.557$, and 
2MASSJ19040934+3639195; $J=11.213$, $J-K=0.703$) would
avoid areas of the array with known bad pixels or significant
gradients in the flat-field, as well as areas known to be affected
by scattered starlight.  We also ensured that a nearby
star (2MASSJ19040089+3639564; $J=5.57$, $J-K=0.941$) that was 
much brighter than the target would not fall in one of several 
regions outside the FOV that are known to 
result in significant scattered light on the detectors.

We observed the field for 5.6~hours spanning UT 30-31 October 2004.
We obtained 1518 full-array images in each of the two bandpasses;
the cadence was 13.2~s with an effective integration
time of 10.4~s.  The images were bundled in sets of 200,
corresponding to the maximum number of images in a single Astronomical 
Observing Request (AOR).  As the starting point in the following analysis, 
we used the IRAC Basic Calibrated Data (BCD) frames.
These frames are produced by the standard IRAC calibration pipeline, 
and include corrections for dark current, flat-fielding, detector
non-linearity, and conversion to flux units.
We converted the time stamps in the image headers to Julian dates and
adjusted these to correspond to the center of the integration.
We then corrected these dates for the light travel time across the 
solar system, including the correction for the spatial
separation of {\it Spitzer} and the Earth (an adjustment of approximately $+29.7$~s).

We evaluated the arithmetic centroid of the target and both 
calibrators in each image.  We found that the pointing jitter was
less than 0.05 pixels ($0\farcs06$) over the course of a single AOR, 
but offsets as large as 0.2 pixels ($0\farcs24$) occurred between AORs.
These offsets were likely due to the reacquisiton
sequence that occurred automatically at the start of each AOR.  
Since these sub-pixel offsets introduce apparent variations
in the recorded flux (clearly evident in our 4.5~$\mu$m photometry, 
as described below), deactivating this 
process would enable even greater photometric precision in future applications
requiring multiple AORs. 
 
We then performed aperture photometry on the images, using an aperture 
radius of 4.0 pixels and a sky estimate derived from an annulus of 
15 to 25 pixels.  (Note that we retained the use of relative fluxes, as 
opposed to magnitudes, throughout the analysis.)
The aperture radius was selected to
minimize the root-mean-square (RMS) residual of the resulting 
time series for the calibrators:  Apertures smaller than 3.5~pixels showed
large deviations due to flux spilling out of the aperture,
whereas radii larger than 4.5 pixels resulted in significantly degraded
RMS residual due to increased sky background (particularly for the 8.0~$\mu$m channel, 
where the sky is brighter).  We also rejected the data from the
first of the 8 AORs, since these data showed increased noise
that presumably resulted from the instrument reaching a new equilibrium
at the settings specific to this campaign.  (We performed aperture photometry on 
several dozen fainter stars in the FOV.  Since many of the resulting time series 
showed a similar effect, we concluded that the omission of these data was justified.)  
We had planned for this possibility by centering the predicted time of secondary 
eclipse within the span of the last 7 AORs.  As a result, the first AOR occurred
well outside the predicted duration of the secondary eclipse, and the 
omission of these data did not impact our conclusions.  We also rejected very 
large single-point deviants (presumably radiation events), which removed a further
2.3\% of the time-series data.

The 8.0~$\mu$m time series for all three stars 
show a increasing trend in brightness over the duration
of the observations, with a rise of roughly 1.5\% from
start to finish. We fit a 3rd-order polynomial to each of
the two calibrators, calculated the average of these
fits, and divided the target time series by this average.
We then normalized the time series by evaluating the average
of those time series data far from the predicted time
of eclipse.  The RMS residual of the resulting normalized time series 
(as evaluated outside times of secondary eclipse) was 0.0085.  We then 
binned the data into 40 bins, so that each bin spanned 7.0~min and 
typically contained 32 individual measurements.  The resulting binned 
time series is shown in Fig.~1.  The assigned error bars in that
figure are the RMS residual of the unbinned data divided by the
square-root of the number of data points in that bin.

The 4.5~$\mu$m data showed no overall brightness trend, in contrast to what 
was seen for the 8.0~$\mu$m data.  These data did, however, show 
photometric variations with a typical amplitude of 0.5\% that clearly 
correlated with centroid position.  
There are several possible explanations for this effect:
(1) The flat-field (which is constructed from images of the sky over
regions of high zodiacal light emission) is less precise at
these shorter wavelengths, since the zodiacal emission is less; 
(2) The 4.5~$\mu$m InSb array may possess intra-pixel sensitivity 
variations, and this effect is more acute at these shorter wavelengths 
due to the greater degree of undersampling; (3) The 4.5~$\mu$m observations
are exposed to greater relative well-depth, and may
enter the regime where the pipeline linearity correction may not 
be adequate, resulting in apparent photometric variations that
correlate with concentration of light in the central brightest
pixel.  We modeled each of these scenarios in detail, and found
that each (or a combination of several) could plausibly explain the photometric 
variations.  The final approach we adopted was simply to decorrelate the 
photometric time series of the target versus X and Y position,
and then quantify any potential attenuation in the signal through
the modeling procedure described in \S3.
As an additional test, we applied this decorrelation procedure to one of the two other
bright stars in the field of view (the other available calibrator was
nearing saturation in these images), and it resulted in a time series 
that showed no significant deviations from a constant brightness. 
After decorrelation, the RMS residual of the normalized time series for the 
target star was 0.0027, a factor of 3.1 smaller
than the 8.0~$\mu$m data (owing to the greater number of photons recorded
at these shorter wavelengths).  We then binned the time series and assigned
error bars in the same fashion as the 8.0~$\mu$m data.  The final binned time series 
is shown in Fig.~2.

\section{Analysis of Time Series}
In order to evaluate the statistical significance of the signal, and to estimate
its amplitude and timing, we created a model eclipse curve by assuming
the system parameters as derived by \citet{sozzetti2004}:
$R_{*} = 0.83 \ R_{\Sun}$, $R_{p} = 1.04 \ R_{\rm Jup}$, $i = 89{\fdg}5$, 
$a = 0.0394$~AU.  We calculated the predicted time of center of secondary eclipse 
to be $t^{\rm predicted}_{\rm II} = 2453309.52363 \pm 0.00052$~HJD, based on the period
and time of center of transit published by \citet{alonso2004}, and the assumption
of a circular orbit ($e=0$).  The uncertainty in this prediction is much
less than that of our estimate of the observed time of secondary eclipse (below),
and hence is not a significant source of error.  We likewise 
neglected the apparent delay in the time of secondary eclipse (relative to the 
times of primary eclipse) resulting from the light travel time across the TrES-1 system, 
$2 a / c = 39.3$~s.
We scaled the model eclipse depth to an amplitude 
$\Delta F_{\rm II}$ and shifted it by a time 
$\Delta t_{\rm II}$, and evaluated the $\chi^{2}$ of the 
resulting fit to the unbinned time series.  
We repeated this procedure over a grid of assumed values for 
$\Delta F_{\rm II}$ and $\Delta t_{\rm II}$.  Throughout this
modeling procedure we analyzed the unbinned time series 
(binned data are shown in Figs.~1 \& 2 for clarity).  

For the 8.0~$\mu$m time series,  the estimated
eclipse depth was $\Delta F_{\rm II} = 0.00225 \pm 0.00036$,
and the timing offset was $\Delta t_{\rm II} = +8.3 \pm 5.2$~min.
The minimum of the chi-squared was $\chi^{2} = 1238.2$ for $(N-2) = 1280$ 
degrees of freedom.  The reduced chi-squared was therefore $\chi^{2}/(N-2) = 0.97$, 
indicating an excellent fit to the data.  
In Fig.~1, we show the binned data overplotted with the best-fit
model, and a contour plot showing the 
corresponding limits on $\Delta F_{\rm II}$ and $\Delta t_{\rm II}$.

For the 4.5~$\mu$m time series, the same procedure initially
yielded an eclipse depth of $\Delta F_{\rm II} = 0.00057 \pm 0.00013$
and a timing offset of $\Delta t_{\rm II} = +19.6 \pm 6.6$~min.
Formally, this offset in the time of secondary eclipse is
statistically very significant.  However, we noted that a small
subset (9.8\%) of the data near the time of ingress (spanning dates 
$-0.05 < t - t^{\rm predicted}_{\rm II} < -0.03$) was entirely responsible
for the displacement in the best-fit value of $\Delta t_{\rm II}$.
Upon closer scrutiny, we found these data to be 
suspicious for two reasons:  First, they originated
from the AOR representing the largest displacement in centroid position,
and hence may not be well-corrected by our decorrelation method.
Second, if these data are valid, the implied duration of secondary eclipse
is too short to be physically plausible (since the egress occurred
at the predicted time).  As a test, we repeated the fitting
procedure, but gave these suspect data zero statistical weight.
The resulting eclipse depth was $\Delta F_{\rm II} = 0.00066 \pm 0.00013$,
consistent with the previous value.  However, the 
timing offset was reduced to $\Delta t_{\rm II} = +8.1 \pm 6.6$~min.
Since we are uncertain whether this discrepancy is in fact
instrumental in nature, we simply treat the difference in the
two estimates of $\Delta t_{\rm II}$ as indicative of the possible level of 
systematic error in the 4.5~$\mu$m eclipse time.
The minimum of the chi-squared was $\chi^{2} = 1220.9$ for $(N-2)=1153$ degrees of freedom. 
The reduced chi-squared was therefore $\chi^{2}/(N-2) = 1.06$, indicating a good fit to the
data.  In Fig.~2, we show the binned data overplotted with both model
fits, and a contour plot showing the corresponding limits on $\Delta F_{\rm II}$ 
and $\Delta t_{\rm II}$ resulting from each analysis.

Of further concern is that the decorrelation procedure
might attenuate the signal in the 4.5~$\mu$m time series.
In order to quantify this effect, we injected eclipses
in the raw time series with depths of $\Delta F_{\rm II} = \{0.0005,0.0010\}$.  
We then subjected these modified time series to the same procedures of decorrelation 
and $\chi^{2}$-fitting.  The derived best-fit eclipse depths were
$\Delta F_{\rm II} = \{0.00110 \pm 0.00013, 0.00155 \pm 0.00013\}$ (respectively), 
indicating that virtually no attenuation had occurred (i.e. the derived signal depths 
were found to be the sum of the original signal and the injected signal).

In summary, we find eclipse depths of 
$\Delta F_{\rm II} = 0.00225 \pm 0.00036$~(8.0~$\mu$m) and 
$\Delta F_{\rm II} = 0.00066 \pm 0.00013$~(4.5~$\mu$m).  Our best estimate
of the time of secondary eclipse is $t_{\rm II} = 2453309.5294 \pm 0.0036$~HJD,
corresponding to a timing offset of $\Delta t_{\rm II} = +8.3 \pm 5.2$~min
from the prediction based on a circular orbit and the period
and time of transit published in \citet{alonso2004}.  Due to the large systematic
uncertainty in the timing estimate based on the 4.5~$\mu$m data, the derived
value of $\Delta t_{\rm II}$ is essentially dictated by the 8.0~$\mu$m data.

\section{Discussion and Conclusions}

\subsection{Planetary Emission}
If we denote the planetary and stellar surface fluxes by
$F_{p}(\lambda)$ and $F_{*}(\lambda)$, respectively, then the predicted signal 
in a given bandpass is prescribed by 
\begin{equation}
\Delta F_{\rm II} \simeq \left( \frac{R_{p}}{R_{*}} \right)^{2} \, \frac{\int F_{p}(\lambda) \ S(\lambda) \ ({\lambda} \, / h \, c) \ d\lambda}{\int F_{*}(\lambda) \ S(\lambda) \ ({\lambda} \, / h \, c) \ d\lambda} 
\end{equation}
where $S(\lambda)$ is the appropriate IRAC spectral response 
function\footnote{IRAC spectral response tables are available at
\texttt{http://ssc.spitzer.caltech.edu/irac/spectral{\_}response.html}} in units of [e$-$/photon].  
In this approach, we neglect the contribution from reflected starlight to the observed values 
of $\Delta F_{\rm II}$, which could be as large as $(R_{p} / a)^{2} \simeq 0.00016$ 
\citep{charbonneau1999} for the TrES-1 system.  The ratio
$(R_{p}/R_{*})^2$ in Eq.~(1) is very accurately constrained by transit
observations, and hence contributes little uncertainty.  In the following discussion, we
assume a model for the stellar flux $F_{*}(\lambda)$ calculated\footnote{The model
for the stellar flux as calculated by R.~Kurucz is available at \texttt{http://kurucz.harvard.edu/stars/K0V/}} 
by R. Kurucz \citep[personal communication, 2005;][]{kurucz1992, kurucz1993} 
for the stellar parameters as derived in \cite{sozzetti2004}.
We note that a Planck curve should not be assumed for the stellar flux in Eq.~(1),
since the stellar flux at infrared wavelengths is significantly less than the prediction 
from a blackbody of the stellar effective temperature.

Using Eq.~(1), we can convert the eclipse depths $\Delta F_{\rm II}$ into 
band-dependent brightness temperatures $T_{4.5\mu{\rm m}}$ and $T_{8.0\mu{\rm m}}$
by substituting $F_{p}(\lambda) = \pi \, B_{\lambda} (T)$, where $B_{\lambda}(T)$
denotes the Planck curve.  Minimizing the value of ${\chi}^2(T)$ for each
of the two bands, we estimate brightness temperatures of $T_{4.5\mu{\rm m}} = 1010 \pm 60$~K 
and $T_{8.0\mu{\rm m}} = 1230 \pm 110$~K.  Based on the modest discrepancy between these estimates
of the brightness temperature, we find that the planetary emission is somewhat inconsistent
with that of a blackbody.  For the purposes of discussion, we initially consider 
the scenario in which the planet is assumed to emit as a blackbody;
we then re-evaluate our results in the light of a more sophisticated model
of the planetary emission.

Treating the planetary emission as a Planck curve of temperature $T_{p}$, we modify the 
procedure described in the previous paragraph, minimizing the summed value of $\chi^{2}(T_{p})$ 
for the two measurements of $\Delta F_{\rm II}$.  In this manner, we estimate the planetary 
temperature to be $T_{p} = 1060 \pm 50$~K.  Under the further assumption that the 
planet is in thermal equilibrium with the stellar radiation, the planetary equilibrium
temperature is prescribed by 
\begin{equation}
T_{\rm eq} = T_{*} \ (R_{*} / 2 \, a)^{1/2} \ [f \, (1-A)]^{1/4} \simeq 1163 \, [f \, (1-A)]^{1/4} \ {\rm K}, 
\end{equation}
where $A$ is the wavelength-integrated Bond albedo (the fraction of energy re-emitted 
relative to the amount received), and the factor $f$ is 1 if the planetary emission
is isotropic, and 2 if only the dayside of the planet re-radiates the
absorbed heat.  We note that this approach neglects the contribution
from the intrinsic luminosity of the planet, which results from ongoing contraction.
Substituting our temperature estimate $T_{p}$ for $T_{\rm eq}$ in Eq.~(2), 
we find $[f \, (1-A)] \simeq 0.69 \pm 0.14$.  If we further assume that the
emission is isotropic ($f = 1$), this implies that the Bond albedo
is $A \simeq 0.31 \pm 0.14$.  The conclusion would be that 
the planet likely absorbs the majority of radiation incident upon it, a 
realization of significant impact to models of hot Jupiter atmospheres
\citep[see][and references therein]{burrows2005, cho2003, marley1999, 
seager2000, sudarsky2003}.  A low albedo
would also be consistent with the upper limits reported by searches for reflected 
light from other hot Jupiters 
\citep{charbonneau1999, cameron2002, leigh2003}, although we
note that those studies constrained a combination of the wavelength-dependent
geometric albedo and phase function, whereas the results presented
here serve to constrain the wavelength-integrated and phase-integrated
quantity $A$.  We reiterate that this discussion assumes isotropic emission;
if $f>1$, then larger values of the Bond albedo $A$ would be permitted.

As noted above, the assumption of blackbody emission is not supported
by the discrepancy (albeit a modest one) in the brightness temperature estimates. 
In particular, the emission at 4.5~$\mu$m is less than,
and the emission at 8.0~$\mu$m is larger than that predicted for a 
blackbody of $T_{p} = 1060$~K.  Numerous theoretical
models of the emission spectra of hot Jupiters have been presented
in the literature (see Burrows 2005, and references therein).
Despite the variety of published models, the consensus is that large 
deviations from the Planck curve are predicted.  In order to illustrate such deviations, 
we consider a model\footnote{This model is publicly available at 
\texttt{http://zenith.as.arizona.edu/$\sim$burrows/sbh/sbh.html}} for the planetary flux 
presented by \cite{sudarsky2003}.  We select the model for 
51~Peg~b, since, of the available choices, that planet most resembles TrES-1
in mass and degree of stellar insolation.  
We divide the model fluxes by those of the stellar model described above;
the resulting prediction for the wavelength-dependent flux ratio is
shown in Fig.~3, overplotted with the two available
data points.  It should be noted that no parameters
have been adjusted to improve the model fit, and, in particular, the model has not been
re-normalized.  Although the model prediction is roughly consistent with the observed 
value at 8.0~$\mu$m, the model over-predicts the flux ratio at 4.5~$\mu$m.  
The emergent flux in the spectral region spanned by the IRAC bandpasses is dominated by the
presence of water and carbon monoxide; methane is not expected to be spectroscopically
prominent since CO is the preferred repository for carbon at these warm
atmospheric temperatures.  The predicted emission peak
near 4~$\mu$m results from a window of low opacity that is constrained
by significant water absorption bands blueward of the peak, and absorption
bands of both water and CO redward of it.  An additional source
of opacity in this region could suppress the flux to the low level that is observed; 
otherwise, it may be that the CO (1$-$0) fundamental band that overlaps 
the red half of the 4.5~$\mu$m bandpass is more prominent than
the model predicts.  \cite{deming2005a}
have recently presented a stringent upper limit on the spectroscopic signature
of CO for HD~209458b.  That study utilized the technique of transmission spectroscopy
(as opposed to emission), and examined a different band, the (2$-$0) transition at 2.3~$\mu$m.
A detailed comparison of theoretical atmospheric models to both the
results presented here and that earlier study would be fruitful.

We note that {\it Spitzer} observations of TrES-1 in the
two additional IRAC bands (3.6~$\mu$m and 5.8~$\mu$m) would likely prove diagnostic of the 
planetary atmosphere.  In particular, the model shown in Fig.~3 predicts a
large excess of emission in the 3.6~$\mu$m bandpass over the prediction of the best-fit blackbody spectrum.  
IRAC observations of HD~209458 would be facilitated by the greater apparent brightness of that system; however, 
detector saturation limits will require the use of a subarray mode, which
may degrade the photometric stability.  If in fact the IRAC subarray mode is as photometrically
stable as the full-array mode, an additional scientific
opportunity will be enabled:  By gathering high-cadence photometric observations during
times of ingress and egress of secondary eclipse, it may be possible to discern
differences in the shape of the lightcurve from the prediction of a uniformly-illuminated 
disk.  Such observations would thus serve to resolve {\it spatially} the planetary
emission across the dayside portion of the planetary disk.  Moreover, the variations observed during ingress 
and egress will be complementary; toward the end of ingress, only the trailing edge of the
planet is unocculted, whereas during the initial portion of egress, only the leading edge of the planet 
is revealed.  Detailed dynamical models of this planet's atmosphere 
\citep{cho2003, menou2003, cooper2005} predict significant atmospheric flows in response to the 
large incident stellar radiation on the dayside.  In particular, some models predict the presence of a hot
spot, shifted downstream from the sub-solar point due to the equatorial atmospheric
flow.  One effect of such a feature would be to shift the apparent center of the
time of secondary eclipse.  As a result, an apparent offset in the time of
secondary eclipse from the prediction of a circular orbit may be induced even if
the eccentricity is zero.  Although such an offset is not detected for the TrES-1 system with
high significance, we encourage theoretical calculations to evaluate whether
apparent offsets of $\sim$8 minutes (i.e. $\sim$1/4 of the planetary disk crossing time) are plausible.

\subsection{Orbital Eccentricity}
The detection of the secondary eclipse is also of interest
because it places constraints on the orbital eccentricity in
addition to those provided by the radial-velocity observations.
A non-zero value of the orbital eccentricity could produce a measurable
shift in the separation of the times of center of primary ($t_{\rm I}$) and
secondary ($t_{\rm II}$) eclipse away from a half-period.  The approximate
formula for the offset \citep[see][]{kalrath1998, charbonneau2003} is 
\begin{equation}
\frac{\pi}{2 \, P} \left( t_{\rm I} - t_{\rm II} - \frac{P}{2} \right) \simeq e \, \cos{\omega}, 
\end{equation}
where $P$ is the orbital period and $\omega$ is the longitude of periastron; the
expression in parentheses is precisely the offset $\Delta t_{\rm II}$ we have estimated
earlier.  Thus, we find that $e$ and $\omega$ must satisfy 
\begin{equation}
e \, \cos{\omega} \simeq \frac{\pi \, \Delta t_{\rm II}}{2 \, P} \simeq 0.0030 \pm 0.0019. 
\end{equation}
From this equation, a value of $e$ may be estimated if an independent
measure of $\omega$ exists.  In particular, we note that an eccentric orbit also 
affects the relative durations of the primary ($\Theta_{\rm I}$) and secondary ($\Theta_{\rm II}$) 
eclipses according to:
\begin{equation}
\frac{\Theta_{\rm I} - \Theta_{\rm II}}{\Theta_{\rm I} + \Theta_{\rm II}} \simeq e \, \sin{\omega}.
\end{equation}
Thus in principle, the separate dependencies of Eqs.~(3) \& (5) on $\omega$ 
permits a direct evaluation of $e$.  In practice, however, since the
size of the second effect is proportional to only the duration of the
eclipse [as opposed to the orbital period, Eq.~(3)], the relative uncertainty
that results in estimates of the duration of the secondary eclipse
is too great to be useful.

The most robust limits on $e$ are obtained by using all available
information on the geometry of the orbit, namely, the times of transit,
the radial velocities, and the time of secondary eclipse reported here.
Constraints on the eccentricity come mainly
from the radial velocities and the time of secondary eclipse. The individual
transit observations (listed in Table~1) are described 
by \cite{alonso2004}, with the exception of event number 11, which
was observed at the IAC~80~cm telescope after that paper went to press.
Of the twelve transit timings available, we excluded the one from the University of Colorado 
Sommers-Bausch Observatory (SBO), as it is discrepant by more than 6-$\sigma$;
we also note that anomalies in the ingress and egress of that event 
are apparent in Fig.~1 of \cite{alonso2004}.  For the radial
velocities, in addition to those reported by \cite{alonso2004}, we
used the measurements by \cite{laughlin2005}, which are of
higher internal precision and provide complementary phase coverage,
particularly near the time of secondary eclipse.

We mapped the resulting $\chi^2$ surface by performing orbital
solutions for fixed values of $e$ and $\omega$ over a 
densely sampled grid of these values.  In each case we
used the timings and radial velocities simultaneously, and solved for
the period and time of transit, as well as the velocity
semi-amplitude and a velocity offset between the measurements by
\cite{laughlin2005} and those of \cite{alonso2004}.  In Table~1, we list the
Observed$-$Calculated ($O-C$) times, and relative errors $(O-C)/\sigma_{\rm HJD}$,
where $\sigma_{\rm HJD}$ are the uncertainties in the timing measurements. 
We list these values both for the best-fit orbit where we forced $e=0$, and for the 
solution where we allowed $e$ and $\omega$ to vary (in which case
the best-fit values were $e = 0.041$ and $\omega = 274\fdg4$; see
below).  
We present these 
times as they will likely find a variety of additional applications,
such as the search for additional planets \citep{agol2004, holman2004}.  
We note that allowing $e$ and $\omega$ to vary noticeably 
improved the fit to the time of secondary eclipse (see Table~1),
reducing the value of $(O-C)/\sigma_{\rm HJD}$ from $+1.53$ to $-0.15$.

A contour plot of the $\chi^{2}(e,\omega)$ surface, and the 
corresponding limits on $e$ and $\omega$, are shown in Fig.~4. 
The minimum of this surface occurs at $e = 0.041$ and $\omega = 274\fdg4$.
The allowed $\{e,\omega\}$ parameter space is very small
yet highly correlated.  The eccentricity is consistent with zero,
however a value as large as $e=0.06$ is permitted
for a small range of values of $\cos{\omega}$ near 0:
In particular, $e>0.02$ requires $270^{\degr}<\omega<285^{\degr}$,
which would mean the orbital ellipse is nearly aligned with our
line of sight.  
Additional radial velocity measurements at key portions of the orbital
phase will serve to further restrict the allowed range of parameter
space.

\cite{bodenheimer2001, bodenheimer2003} and \cite{laughlin2005} have proposed that 
the tidal dissipation of a nonzero orbital eccentricity could generate
an internal energy source sufficient to increase significantly the planetary radius.
The precise value of the planetary radius could be used to
infer the presence (or absence) of a rocky core, which would be a
critical clue to the formation process of this planet.  However, this
effect is masked if an additional source of energy serves
to preserve an inflated value for the planetary radius \citep{laughlin2005}.
Hence it is important for us to evaluate the size of this energy
term for TrES-1, even if its radius is already roughly consistent with
model expectations.

The circularization timescale $\tau_{\rm circ}$ for TrES-1 is given
by \citep{goldreich1966}:
\begin{equation}
\tau_{\rm circ} = \frac{e}{\dot{e}} = \left( \frac{2 \, P \, Q_{p}}{63 \, \pi}\right) 
\left( \frac{M_{p}}{M_{*}} \right) \left( \frac{a}{R_{p}} \right)^{5} \simeq 
0.21 \left( \frac{Q_{p}}{10^{6}} \right) {\rm Gyr},
\end{equation}
where $Q_{p}$ is the tidal quality factor.  There is a large uncertainty
in the likely value of $Q_{p}$.  \cite{terquem1998} find that
$Q$ is of order $10^{6}$ for stars (based on the tidal circularization
of main-sequence binaries), and \cite{yoder1981} constrain the
value for Jupiter to the range $6 \times 10^{4} < Q_{\rm Jup} < 2 \times 10^{6}$ 
(based on the Jupiter-Io interaction).
For an assumed value of $Q_{p}$, the timescale in Eq.~(6) can then be used
to estimate the rate of energy dissipation in the planet \citep{bodenheimer2003}:
\begin{equation}
\dot{E}_d = \frac{e^{2} \, G \, M_{*} \, M_{p}}{a \, \tau_{\rm circ}} \simeq 
4.3 \times 10^{28} \, e^{2} \left( \frac{10^{6}}{Q_{p}} \right) \, {\rm erg \, s^{-1}}. 
\end{equation}
If TrES-1 has no solid core, a value of 
$\dot{E}_{d} \lesssim 1 \times 10^{26} {\rm \ erg \, s^{-1}}$ \citep[][also G.~Laughlin, personal communication, 2005]{laughlin2005}
is required; otherwise, the observed radius is too small.  Using Eq.~(7), we find that this in turn
requires $e \lesssim 0.048 \, (Q_{p}/10^{6})^{1/2}$, which indeed
appears consistent with our findings.  We encourage additional radial velocity
observations at key portions of the orbital phase:  When combined
with the times of primary and secondary eclipse presented in Table~1, such data
will shrink the allowed range of values for the eccentricity and,
by implication, further constrain the rate of energy dissipation.

This discussion serves to motivate
similar work for HD~209458b.  In contrast to TrES-1, its radius requires 
an internal source of energy.  The detection of
of a significantly non-zero eccentricity for HD~209458b would 
provide strong support for this scenario:
Of the various models for the inflated radius of a hot Jupiter \citep[e.g.][]{bodenheimer2001,
guillot2002, burrows2003}, the tidal dissipation of orbital eccentricity most
naturally leads to a variety of observed planetary radii, since
only some systems may contain additional planets that serve to pump
the orbital eccentricity.

{\it Shortly after we submitted this manuscript, we learned of
a similar detection by \cite{deming2005b}, which presents Spitzer/MIPS
24~$\mu$m photometry spanning a time of secondary eclipse of the 
HD~209458 planetary system.  Additional Spitzer observations of these
and other extrasolar planets will enable the first comparative studies
of the thermal emission from these elusive objects.}

\acknowledgments
This work is based on observations made with the {\it Spitzer Space Telescope}, 
which is operated by the Jet Propulsion Laboratory, California Institute of Technology 
under NASA contract 1407.  G.T. acknowledges partial support for
this work from NASA Origins grant NNG04LG89G.  R.A. acknowledges financial support from 
grants AyA2001-1571 and ESP2001-4529-PE of the Spanish National Research plan.
We are grateful to Robert Kurucz for providing the model of the stellar flux.
We thank Joseph Hora and John Stauffer for illuminating IRAC discussions, and
we thank Drake Deming, Scott Gaudi, Robert Noyes, Dimitar Sasselov, Sara Seager,
and Joshua Winn for comments that improved the manuscript.

Facilities: \facility{Spitzer(IRAC)}.

\clearpage

\begin{deluxetable}{llcrccccc}
\rotate
\tablecaption{Times of Primary and Secondary Eclipse}
\tablewidth{0pt}
\tablehead{
\colhead{Event} & \colhead{Type} & \colhead{HJD} & \colhead{$N_{\rm elapsed}$} & \colhead{$\sigma_{\rm HJD}$} & \colhead{$(O-C)^{\rm a}$} & \colhead{$\frac{(O-C)^{\rm a}}{\sigma_{\rm HJD}}$} & \colhead{$(O-C)^{\rm b}$} & \colhead{$\frac{(O-C)^{\rm b}}{\sigma_{\rm HJD}}$}}
\startdata
    1  &  primary  & 2452847.4363 & $-$112.0 &  0.0015 &   $-$0.0022  &  $-$1.45 & $-$0.0023 & $-$1.51\\
    2  &  primary  & 2452850.4709 & $-$111.0 &  0.0016 &   $+$0.0023  &  $+$1.48 & $+$0.0022 & $+$1.42\\
    3  &  primary  & 2452856.5286 & $-$109.0 &  0.0015 &   $-$0.0002  &  $-$0.12 & $-$0.0003 & $-$0.18\\
    4  &  primary  & 2452868.6503 & $-$105.0 &  0.0022 &   $+$0.0013  &  $+$0.59 & $+$0.0012 & $+$0.55\\
    5  &  primary  & 2453171.6523 & $-$5.0   &  0.0019 &   $-$0.0035  &  $-$1.81 & $-$0.0035 & $-$1.80\\
    6  &  primary  & 2453174.6864 & $-$4.0   &  0.0004 &   $+$0.0006  &  $+$1.27 & $+$0.0006 & $+$1.30\\
    *  &  primary$^{\rm c}$  & 2453180.7529 & $-$2.0   &  0.0010 &   $+$0.0068  &  $+$6.61 & $+$0.0069 & $+$6.89\\ 
    7  &  primary  & 2453183.7752 & $-$1.0   &  0.0005 &   $-$0.0009  &  $-$1.66 & $-$0.0008 & $-$1.63\\
    8  &  primary  & 2453186.8061 & \phantom{$+$}0.0   &  0.0003 &   $-$0.0001  &  $-$0.22 & $-$0.0001 & $-$0.16\\
    9  &  primary  & 2453189.8354 & $+$1.0   &  0.0019 &   $-$0.0008  &  $-$0.44 & $-$0.0008 & $-$0.43\\
    10 &  primary  & 2453192.8694 & $+$2.0   &  0.0015 &   $+$0.0031  &  $+$2.01 & $+$0.0031 & $+$2.03\\
    11 &  primary  & 2453247.4075 & $+$20.0  &  0.0004 &   \phantom{$+$}0.0000  &  $+$0.05 & $+$0.0001 & $+$0.16\\
    12 &  secondary & 2453309.5294 & $+$40.5  &  0.0036 &   $+$0.0055  &  $+$1.53 & $-$0.0005 & $-$0.15\\
\enddata
\tablenotetext{a}{Forced $e=0$.}
\tablenotetext{b}{Allowed $e$ and $\omega$ to float.}
\tablenotetext{c}{This event was excluded from the fit (see text).}
\end{deluxetable}

\clearpage

\figcaption{\emph{Upper panel:} Shown are the binned 8.0~$\mu$m time series.  The best-fit
model eclipse curve has a depth of $\Delta F_{\rm II} = 0.00225$ and a timing
offset of $\Delta t_{\rm II} = +8.3$~minutes, and is plotted as the solid black line.
A model of the same depth but $\Delta t_{\rm II} = 0$ is shown as the
dashed line.  \emph{Lower panel:} Shown are the 1$\sigma$, 2$\sigma$, and 3$\sigma$ confidence
ellipses on the eclipse depth and timing offset.}

\figcaption{\emph{Upper panel:} Shown are the binned 4.5~$\mu$m time series.  The best-fit
model eclipse curve (excluding the data at times $-0.05 < t {\rm (days)} < -0.03$; see text) 
has a depth of $\Delta F_{\rm II} = 0.00066$ and a timing offset of $\Delta t_{\rm II} = +8.1$~minutes, 
and is plotted as the solid black line.  A model of the same depth but $\Delta t_{\rm II} = 0$ is shown as the
dashed line.  The best-fit model when these additional
9.8\% of data are included is shown as the grey line; the estimated
depth is similar, but a significant timing offset is found. 
\emph{Lower panel:} Shown as black ellipses are the 1$\sigma$, 2$\sigma$, and 3$\sigma$ confidence
ellipses on the eclipse depth and timing offset for the restricted data set;
the corresponding ellipses for the complete data set are shown in gray.}

\figcaption{The solid black line shows the \cite{sudarsky2003} model hot-Jupiter spectrum
divided by the stellar model spectrum (see text for details).
The open diamonds show the predicted flux ratios for this model integrated
over the four IRAC bandpasses (which are shown in gray, and renormalized for clarity).  
The observed eclipse depths at 4.5~$\mu$m and 8.0~$\mu$m are overplotted
as black diamonds.  No parameters have been adjusted to the model to improve
the fit.  The dotted line shows the best-fit blackbody spectrum (corresponding to a
temperature of 1060~K), divided by the model stellar spectrum.  
Although the \cite{sudarsky2003} model prediction is roughly consistent
with the observations at 8.0~$\mu$m, the model over-predicts the planetary
flux at 4.5~$\mu$m.  The prediction of a relatively large flux ratio 
at 3.6~$\mu$m should be readily testable with additional IRAC observations.}

\figcaption{Shown are the 1$\sigma$, 2$\sigma$, and 3$\sigma$ confidence spaces
of the orbital eccentricity $e$ and the longitude of periastron $\omega$.  Since
it is the combined expression $e \cos{\omega}$ that is constrained by the
time of secondary eclipse, there exists a small range of $\omega$ (near $\cos{\omega}=0$)
where a significant eccentricity is permitted.  For this situation to occur,
however, the orbital ellipse would need to be very nearly aligned with
our line of sight.  Additional radial velocity observations will further restrict
the allowed parameter space.}

\clearpage

\begin{figure}
\plotone{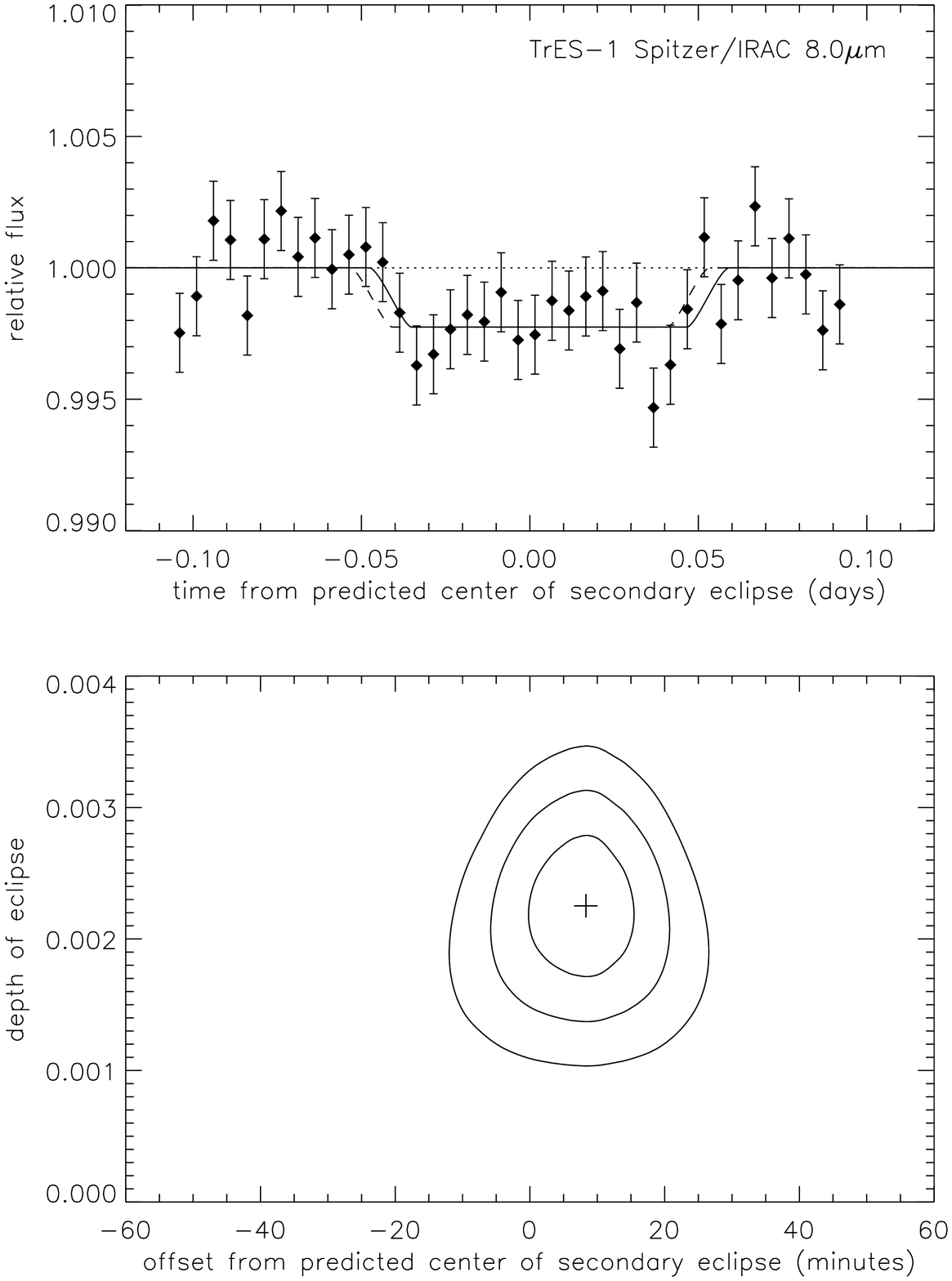}
\end{figure}

\begin{figure}
\plotone{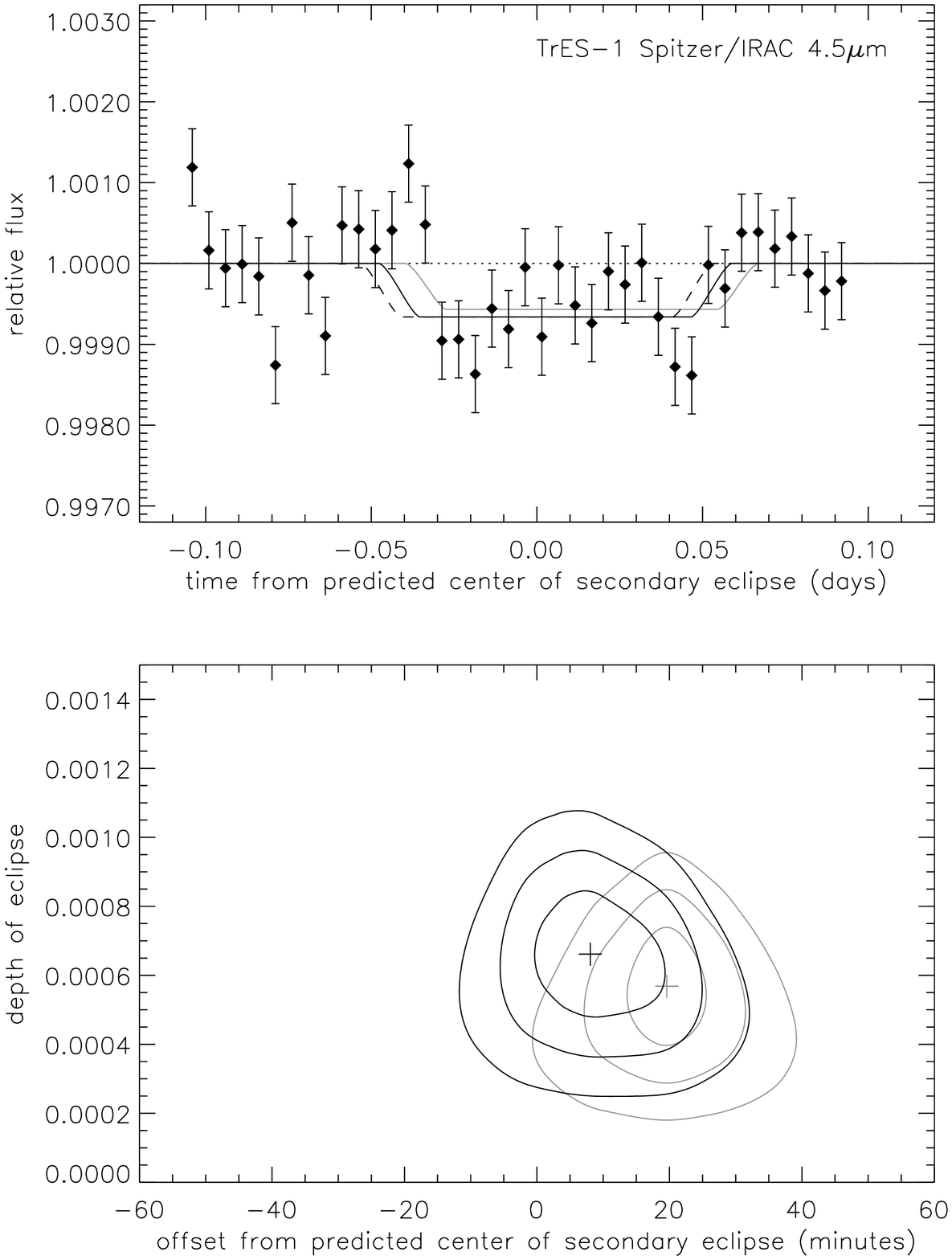}
\end{figure}

\begin{figure}
\plotone{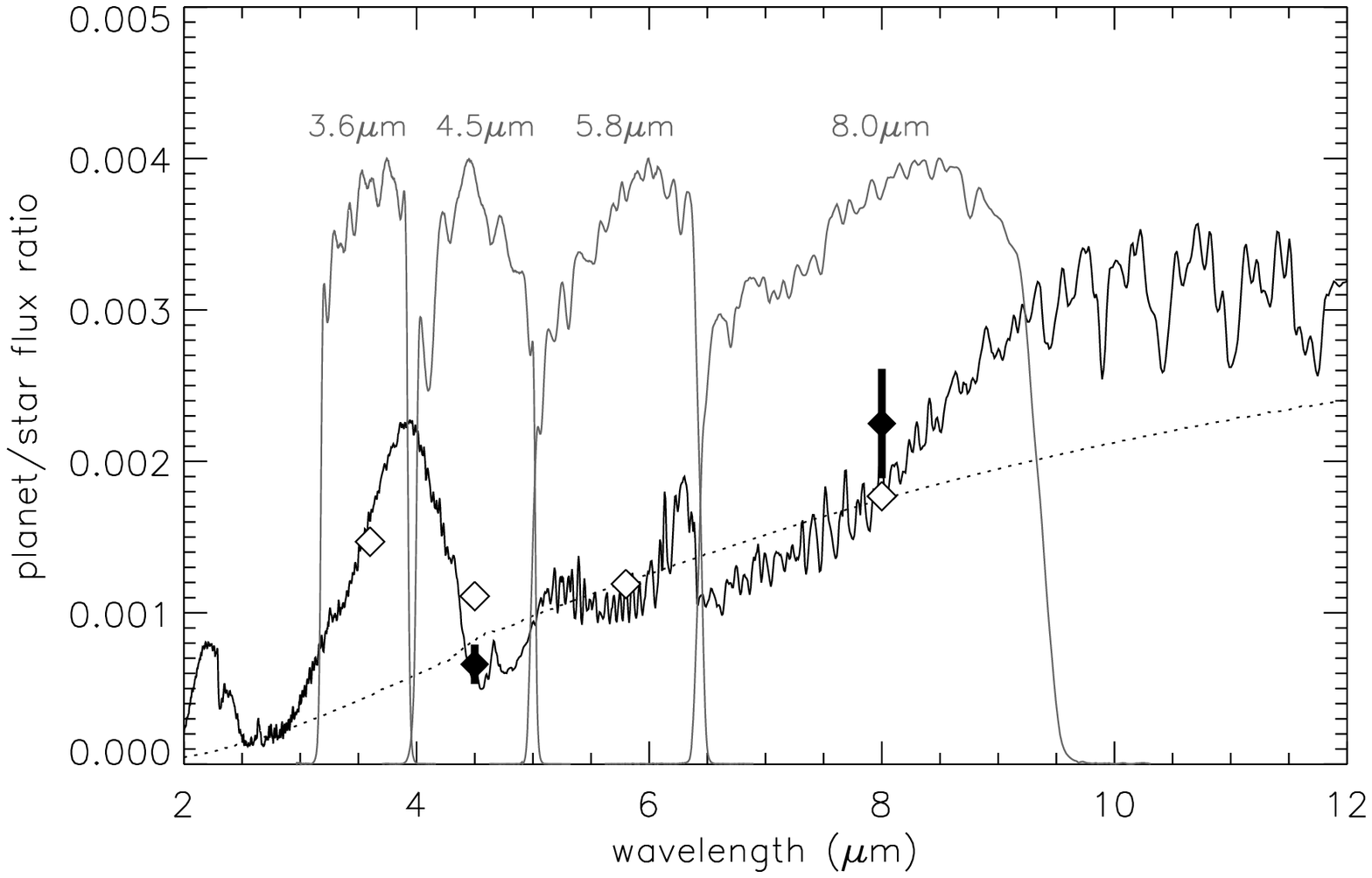}
\end{figure}

\begin{figure}
\plotone{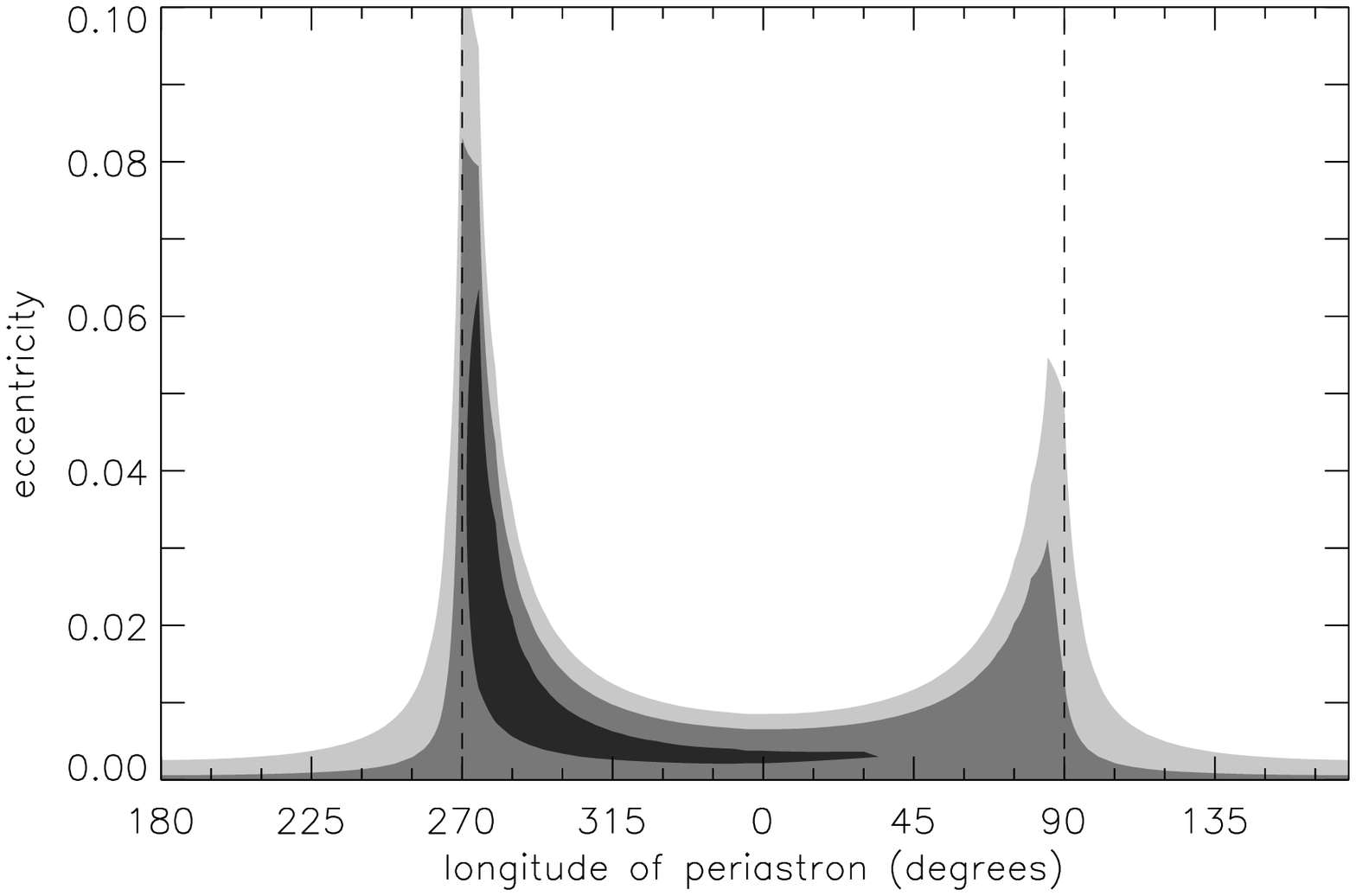}
\end{figure}

\end{document}